\documentclass[
aps,%
12pt,%
final,%
notitlepage,%
oneside,%
onecolumn,%
nobibnotes,%
nofootinbib,
superscriptaddress,%
noshowpacs,%
centertags]%
{revtex4}
\RequirePackage{graphicx}

\def\refitem#1{\relax}

\begin{document}
\title{Search for the QCD Critical Point: Higher Moments of Net-proton Multiplicity Distributions}

\author{\firstname{Xiaofeng} \surname{Luo}}
\email{xfluo@lbl.gov} \affiliation{Department of Modern Physics,
University of Science and Technology of China, Hefei, China}
\affiliation{Lawrence Berkeley National Laboratory Berkeley, CA,USA}

\author{\firstname{B.} \surname{Mohanty}}
\affiliation{Variable Energy Cyclotron Center, Kolkata, India}

\author{\firstname{H.G.} \surname{Ritter}}
\affiliation{Lawrence Berkeley National Laboratory Berkeley, CA,USA}

\author{\firstname{N.} \surname{Xu}}
\affiliation{Lawrence Berkeley National Laboratory Berkeley, CA,USA}

\begin{abstract}
Higher moments of event-by-event net-proton multiplicity
distributions have been applied to search for the QCD critical
point. Model results are used to provide a baseline for this search.
The measured moment products, $\kappa\sigma^2$ and $S\sigma$ of
net-proton distributions, which are directly connected to the
thermodynamical baryon number susceptibility ratio in Lattice QCD
and Hadron Resonance Gas (HRG) model, are compared to the transport
and thermal model results. We argue that a non-monotonic dependence
of $\kappa \sigma^2$ and $S \sigma$ as a function of beam energy can
be used to search for the QCD critical point.
\end{abstract}

\maketitle

\section{Introduction}
Lattice QCD calculations predict that a cross-over from the hadronic
phase to the Quark Gluon Plasma (QGP) phase occurs above a critical
temperature with zero baryon chemical potential ($\mu_{B}$). The
corresponding cross-over temperature range has been estimated to be
about 170 - 190 MeV \cite{aoki}. At large $\mu_{B}$, QCD based model
calculations indicate that the transition from the hadronic phase to
the QGP phase is of first order. The end point of the first order
phase transition line is the QCD Critical Point ( CP )
\cite{ejiri,misha}. There are large theoretical uncertainties of its
location and even its existence is not confirmed~\cite{misha}.
Experimentally, we study the QCD phase diagram by varying the
colliding energy in heavy ion collisions ~\cite{adams}. The
possibility of the existence of the CP has motivated our interest to
search for it with the RHIC beam energy scan program \cite{stephan}.
By tuning the collision energy from a center of mass energy of 200
GeV down to 5 GeV, we will be able to vary the baryon chemical
potential from $\mu_{B} \sim 20$ to $\mu_{B}$ of about 500 MeV.

The characteristic feature of a critical point are a large
correlation length ( $\xi $ ) and critical fluctuations. Recently,
theoretical calculations have shown that higher moments of
multiplicity distributions of conserved quantities, such as
net-baryon, net-charge, and net-strangeness, are sensitive to the
correlation length $\xi$ \cite{stephanov_2}.

In Lattice QCD calculation with $\mu_{B}=0$, higher order
susceptibilities of the baryon number, which can be related to the
higher order moments of the net-baryon multiplicity distributions,
show a non-monotonic behavior near $T_{c}$ \cite{cheng}. A similar
behavior is expected at CP in the finite $\mu_{B}$ region.
Experimentally, it is hard to measure the net-baryon number
event-by-event while the net-proton number is measurable.
Theoretical calculations show that fluctuations of the net-proton
number can be used to infer the net-baryon number fluctuations at
the CP \cite{Hatta}.

In this paper we will show the centrality and energy dependence of
various moments and moment products of net-proton multiplicity
distributions from published STAR Au+Au data~\cite{Data} at
$\sqrt{s_{NN}}=19.6,62.4,200$ GeV and model calculations.

\section{Observables}
The various moment of the event-by-event multiplicity distributions
are defined as: Mean, $M = $ $ <N>$, Variance, $\sigma^{2}$ $=$ $
<(\Delta N)^{2}>$, Skewness, $S={{ < (\Delta N)^3
> }}/{{\sigma ^3 }}$, and Kurtosis, $\kappa={{ < (\Delta N)^4  > }}/{{\sigma ^4 }} - 3
$, where $\Delta N=N-<N>$. Skewness and Kurtosis are used to
characterize the asymmetry and peakness of the multiplicity
distributions, respectively. For gaussian distributions, they are
equal to zero. Thus, the Skewness and Kurtosis are ideal probe to
demonstrate the non-Gaussian fluctuation feature as expected near
the CP. In particular a sign change of the skewness or kurtosis may
be an indication that the system crossed the phase boundary
\cite{stephanov_2,asakawa}.

\section{Results}
We have calculated the various moments of net-proton ($\Delta
p=N_{p}-N_{\bar{p}}$) distributions from transport models ( AMPT
(ver.2.11) \cite{Lin}, Hijing (ver.1.35) \cite{XinNian}, UrQMD
(ver.2.3) \cite{petersen}) and a thermal model (Therminator (ver
1.0) \cite{Kisiel}). By using several models with different physics
implemented, we can study the effects of physics correlations and
backgrounds which are present in the data and that might modify
purely statistical emission patterns, like resonance decays,
jet-production (Hijing), coalescence mechanism of particle
production (AMPT), thermal particle production (Therminator), and
hadronic rescatterring (AMPT,UrQMD).

The kinetic coverage of protons and anti-protons used in our
analysis is $0.4<p_{T}<0.8$ GeV/c and $|y|<0.5$. Fig. 1 shows the
number of participant ($N_{part}$) dependence of four moments ({\it
M}, $\sigma$, {\it S}, $\kappa$) extracted from net-proton
distributions of Au+Au collisions at $\sqrt{s_{NN}}=200$ GeV for the
various models. {\it M} and $\sigma$ show a monotonic increase with
$N_{part}$ for all of the models, while {\it S} and $\kappa$, which
are positive, decrease monotonically with $N_{part}$, which means
the shape of the net-proton distributions become more symmetric as
the centrality increases. The dashed lines in Fig.1 are derived from
the independent emission source model ~\cite{SQM2009}. The
centrality evolution of the various moments of net-proton
distributions in Fig. 1 can be well described by such a model.

Fig. 2 shows the $N_{part}$ dependence of moment products $S
\sigma$, $\kappa \sigma^{2}$ of net-proton distributions from the
AMPT string melting model with parton cross section
$\sigma_{pp}=10mb$ for Au+Au collisions at $\sqrt{s_{NN}}=7.7, 11.5,
19.6, 39, 62.4 $ and $ 200$ GeV. In the upper panel, the $S \sigma$
shows almost no centrality dependence for high energy, but it shows
a small decreasing trend for low energies. We can also see that the
$S \sigma$ has a strong energy dependence decreasing with increasing
energy. In the lower panel of Fig. 2, the $\kappa \sigma^{2}$ shows
almost no centrality dependence and the values are around unity for
all energies.

In Fig. 3, the energy dependence of moment products $S \sigma$,
$\kappa \sigma^{2}$ for most central net-proton distributions from
STAR data~\cite{Data} are compared with the results from various
models. We see the data are in good agreement with the HRG model
($\kappa_B \sigma_B^2=1$, $S_B \sigma_B=tanh(\mu_B/T)$) ~\cite{HRG}
and the thermal model (Therminator) results. HIJING, UrQMD and AMPT
default fail to describe $S \sigma$ and $\kappa \sigma^{2}$
simultaneously. For $\kappa \sigma^{2}$, the results from various
models show no dependence on energy and are close to unity. A large
deviation from constant as a function of $N_{part}$ and collision
energy for $\kappa \sigma^2$ may indicate new physics, such as
critical fluctuations \cite{stephanov_2}. Recent lattice QCD
calculations from \cite{Gupta} have shown that $\kappa \sigma^2$
non-monotonically depends on colliding energy in the neighbourhood
of the critical point.

\section{Summary}

Higher moments of the distributions of conserved quantities are
predicted to be sensitive to the correlation length at CP and to be
related to the susceptibilities computed in Lattice QCD and the HRG
model. Various non-CP models (AMPT, Hijing, Therminator, UrQMD, HRG)
have been applied to study the non-CP physics background effects on
the higher moments of net-proton distributions. The moment products
$S \sigma$, $\kappa \sigma^2$ of net-proton distributions have
almost no dependence on collision centrality and $\kappa \sigma^2$
is also found to be constant as a function of energy for various
models. On the other hand, the high energy data are in good
agreement with the HRG model.

Both the model as well as the data do not show non-monotonic
behavior. Thus they can serve as a baseline of the behavior expected
from known physics effects for the RHIC beam energy scan. The
presence of a critical point in that region may result in
non-gaussian fluctuations and non-monotonic behavior of the
observables studied here as a function of collision energy.

\begin{acknowledgments}
This work was supported in part by the U.S. Department of Energy
under Contract No. DE-AC03-76SF00098 and National Natural Foundation
of China under Grant No. (10835005,10979003) and Major Basic
Research Development Program(2008CB817702). BM is supported by
DAE-BRNS project Sanction No. 2010/21/15-BRNS/2026.

\end{acknowledgments}

\section{References}

\bibliography{CPOD2010}
\bibliographystyle{unsrt}

\newpage
\begin{figure}[ht]
\centering \vspace{0pt}
    \includegraphics[scale=0.4]{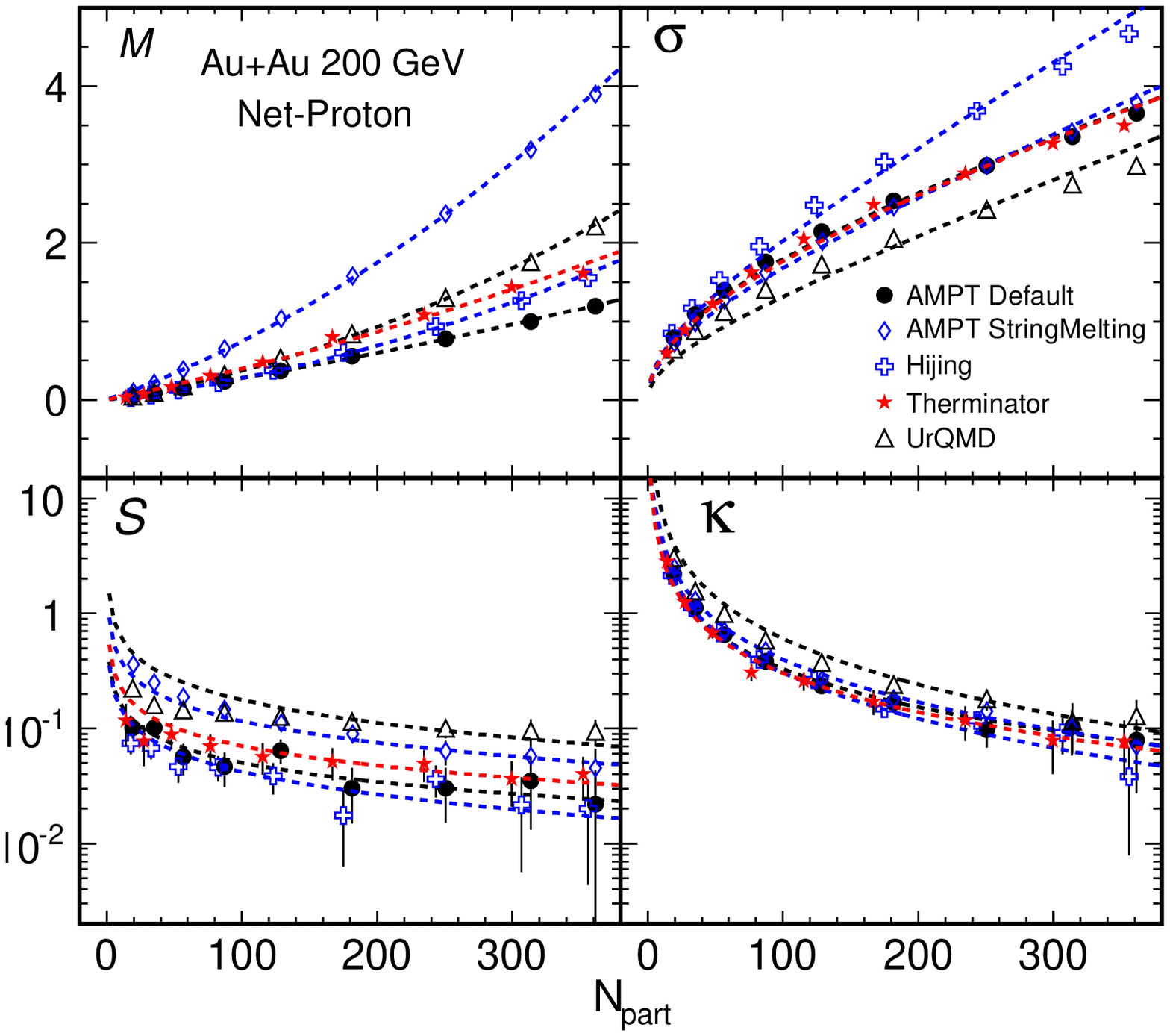}
    \caption{Centrality dependence of various moments of net-proton distributions
for Au+Au collisions at $\sqrt{s_{NN}}=200$ GeV from various models.
The dashed lines represent the expectations for statistical
emission.} \vspace{-0.3in}

    \label{fig:side:a}
  \end{figure}

\begin{figure}[ht]
\centering \vspace{0pt}
    \includegraphics[scale=0.35]{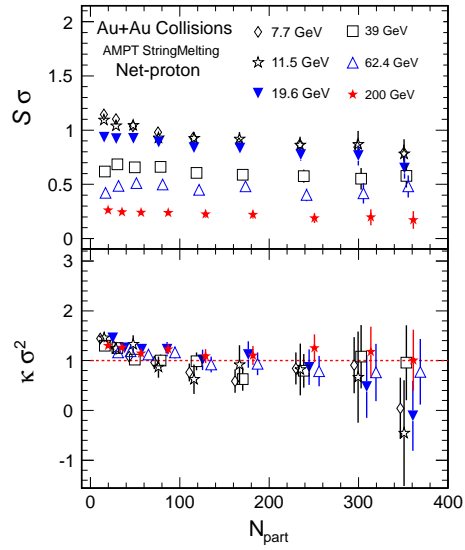}
\vspace{-0.3in}
   \caption{Centrality dependence of moment products $S \sigma$ and $\kappa
\sigma^2$ of net-proton distributions for Au+Au collisions of
various energies from AMPT String Melting model calculation.}
   \label{fig:side:a}
\end{figure}

\begin{figure}[ht]
\centering \vspace{0pt}
    \includegraphics[scale=0.35]{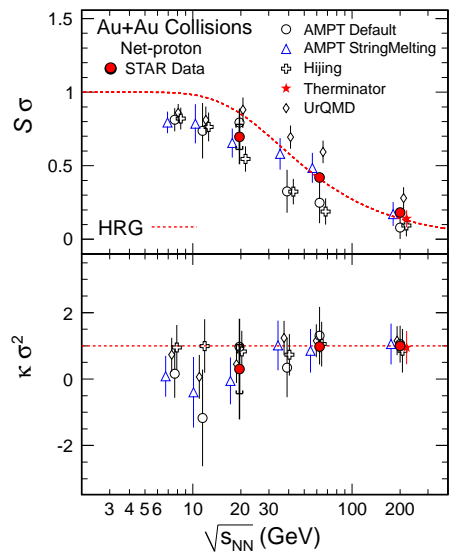}
\vspace{-0.3in}
   \caption{Energy dependence of  moment products $S \sigma$ and $\kappa
\sigma^2$ of net-proton distributions for Au+Au collisions of
various models and STAR data.}
    \label{fig:side:a}
  \end{figure}

\newpage

\begin{center}
FIGURE CAPTIONS
\end{center}
\begin{enumerate}
\item
Centrality dependence of various moments of net-proton distributions
for Au+Au collisions at $\sqrt{s_{NN}}=200$ GeV from various models.
The dashed lines represent the expectations for statistical
emission.
\item
Centrality dependence of moment products $S \sigma$ and $\kappa
\sigma^2$ of net-proton distributions for Au+Au collisions of
various energies from AMPT String Melting model calculation.
\item
Energy dependence of  moment products $S \sigma$ and $\kappa
\sigma^2$ of net-proton distributions for Au+Au collisions of
various models and STAR data.
\end{enumerate}

\end{document}